\documentclass[preprint,times,5p]{elsarticle}

\usepackage{graphicx}
\usepackage{amssymb}
\usepackage{amsmath}

\usepackage{siunitx}
\usepackage{hyperref}
\hypersetup{pdfauthor=author}
\usepackage{booktabs}

\journal{Nucl. Instrum. Methods Phys. Res. A}

\begin{document}

\begin{frontmatter}


\title{Characterization of electroless nickel-phosphorus plating for ultracold-neutron storage}

\author[Nagoya]{H.~Akatsuka}
\author[UM]{T.~Andalib}
\author[TRIUMF]{B.~Bell}
\author[TRIUMF]{J.~Berean-Dutcher}
\author[TRIUMF,UBC]{N.~Bernier}
\author[UW,UM]{C.P.~Bidinosti}
\author[LANL]{C.~Cude-Woods}
\author[LANL]{S.A.~Currie}
\author[TRIUMF]{C.A.~Davis}
\author[TRIUMF,UBC]{B.~Franke}
\author[TRIUMF]{R.~Gaur}
\author[SNOLAB]{P.~Giampa}
\author[UM]{S.~Hansen-Romu}
\author[LANL]{M.T.~Hassan}
\author[RCNP]{K.~Hatanaka}
\author[RCNP]{T.~Higuchi}
\author[TRIUMF]{C.~Gibson}
\author[KEK,JPARC]{G.~Ichikawa}
\author[Nagoya]{I.~Ide}
\author[RCNP]{S.~Imajo}
\author[LANL]{T.M.~Ito}
\author[UM,UW]{B.~Jamieson}
\author[KEK]{S.~Kawasaki}
\author[Nagoya]{M.~Kitaguchi}
\author[UBC]{W.~Klassen}
\author[UNBC]{E.~Korkmaz}
\author[TUM]{F.~Kuchler}
\author[UM,UW]{M.~Lang}
\author[UM]{M.~Lavvaf}
\author[TRIUMF,UW]{T.~Lindner}
\author[LANL]{M.~Makela}
\author[UM]{J.~Mammei}
\author[TRIUMF,UW,UM]{R.~Mammei}
\author[UW,UM]{J.W.~Martin}
\author[TRIUMF,RCNP]{R.~Matsumiya}
\author[UBC]{E.~Miller}
\author[KEK,JPARC]{K.~Mishima}
\author[UBC,TRIUMF]{T.~Momose}
\author[TRIUMF]{S.~Morawetz}
\author[LANL]{C.L.~Morris}
\author[RCNP,IMP-CAS]{H.J.~Ong}
\author[LANL]{C.M.~O'Shaughnessy}
\author[TRIUMF]{M.~Pereira-Wilson}
\author[TRIUMF,SFU]{R.~Picker}
\author[TRIUMF,Coburg]{F.~Piermaier}
\author[TRIUMF,RCNP]{E.~Pierre}
\author[TRIUMF]{W.~Schreyer\corref{correspondingauthor}}
\cortext[correspondingauthor]{Corresponding author}
\ead{wschreyer@triumf.ca}
\author[TRIUMF,SFU]{S.~Sidhu}
\author[TRIUMF,Coburg]{D.~Stang}
\author[TRIUMF]{V.~Tiepo}
\author[UBC,TRIUMF]{S.~Vanbergen}
\author[UBC]{R.~Wang}
\author[LANL,Indiana]{D.~Wong}
\author[Nagoya]{N.~Yamamoto}

\address[TRIUMF]{TRIUMF, Vancouver, BC V6T 2A3, Canada}
\address[UM]{University of Manitoba, Winnipeg, MB R3T 2N2, Canada}
\address[LANL]{Los Alamos National Laboratory, Los Alamos, NM 87545, USA}
\address[UBC]{The University of British Columbia, Vancouver, BC V6T 1Z1, Canada}
\address[UW]{University of Winnipeg, Winnipeg, MB R3B 2E9, Canada}
\address[RCNP]{Research Center for Nuclear Physics, Osaka University, Osaka 567-0047, Japan}
\address[Nagoya]{Nagoya University, Nagoya 464-8601, Japan}
\address[KEK]{KEK, Tsukuba 305-0801, Japan}
\address[SFU]{Simon Fraser University, Burnaby, BC V5A 1S6, Canada}
\address[JPARC]{J-PARC, Tokai 319-1195, Japan}
\address[Coburg]{Coburg University of Applied Science, 96450 Coburg, Germany}
\address[UNBC]{University of Northern British Columbia, Prince George, BC V2N 4Z9, Canada}
\address[SNOLAB]{SNOLAB, Sudbury, ON P3Y 1N2, Canada}
\address[TUM]{Technical University of Munich, 85748 Garching, Germany}
\address[Indiana]{Indiana University, Bloomington, IN 47405, USA}
\address[IMP-CAS]{Institute of Modern Physics, Chinese Academy of Sciences, Lanzhou 730000, China}

\begin{abstract}
Electroless nickel plating is an established industrial process that provides a robust and relatively low-cost coating suitable for transporting and storing ultracold neutrons (UCN). Using roughness measurements and UCN-storage experiments we characterized UCN guides made from polished aluminum or stainless-steel tubes plated by several vendors.

All electroless nickel platings were similarly suited for UCN storage with an average loss probability per wall bounce of \numrange{2.8e-4}{4.1e-4} for energies between \SIlist{90;190}{\nano\electronvolt}, or a ratio of imaginary to real Fermi potential $\eta$ of \numrange{1.7e-4}{3.3e-4}. Measurements at different elevations indicate that the energy dependence of UCN losses is well described by the imaginary Fermi potential. Some special considerations are required to avoid an increase in surface roughness during the plating process and hence a reduction in UCN transmission. Increased roughness had only a minor impact on storage properties.

Based on these findings we chose a vendor to plate the UCN-production vessel that will contain the superfluid-helium converter for the new TRIUMF UltraCold Advanced Neutron (TUCAN) source, achieving acceptable UCN-storage properties with ${\eta=\num{3.5(5)e-4}}$.
\end{abstract}


\end{frontmatter}

\section{Introduction}

Neutrons with energies below a few hundred nano-electronvolts, so-called ultracold neutrons (UCNs), are totally reflected by certain materials under all angles of incidence. Hence, they can be transported to experiments and stored in vessels for several minutes, making them perfectly suitable for precision measurements of the neutron's fundamental properties such as lifetime, beta-decay correlations, electric dipole moment, charge, and interaction with gravity. However, due to the low intensity of available UCN sources, these measurements often take years to gather competitive statistical precision.

To maximize the number of UCNs available to experiments, recent developments of UCN guides have focused on increasing transport efficiency by reducing diffuse neutron reflections, typically by exploiting the very small roughness that can be achieved on glass surfaces. Glass tubes can be coated with materials with high reflective Fermi potential like nickel, nickel alloys, or diamond-like carbon~\cite{Bison2020}. Or the properties of a glass surface can be replicated onto thin nickel-alloy sheets using the ''Replika`` technology~\cite{PLONKA2007450}. The resulting guides are fragile, and seamlessly connecting them without gaps is challenging---gaps and slits often dominate losses in a UCN guide system. For these reasons, polished stainless-steel tubes that can be easily manufactured, cleaned, and connected are still widely in use.

The TRIUMF UltraCold Advanced Neutron (TUCAN) collaboration is building a new UCN source based on a superfluid-helium converter close to a spallation neutron source that is expected to surpass the intensity of existing sources by at least one order of magnitude~\cite{SCHREYER2020163525}. To achieve this performance the converter needs to be contained in a large, thin-walled, superfluid-leak-tight, and radiation-tolerant vessel made from materials with low density and low neutron absorption; ideally an alloy of beryllium, magnesium, or aluminum. Beryllium is light-weight (\SI{1.8}{\gram\per\cubic\centi\meter}), has a small neutron-absorption cross section (\SI{0.008}{\barn}), and has a high Fermi potential (\SI{251}{\nano\electronvolt}), but is prohibitively expensive. Magnesium and aluminum have similar densities (\SIrange{1.7}{2.7}{\gram\per\cubic\centi\meter}) but higher neutron-absorption cross sections (\SIrange{0.06}{0.23}{\barn}) and low Fermi potentials (\SIrange{54}{60}{\nano\electronvolt}), requiring a robust, UCN-reflective coating. We did consider beryllium-aluminum~\cite{AlBeMet} and magnesium-aluminum alloys~\cite{SCHREYER2020163525}, however these uncommon materials would have required specialized knowledge in machining, welding, and plating to fulfill the challenging requirements. We ultimately chose aluminum due to the large variety of available alloys and wealth of knowledge to complete the vessel within a reasonable time and budget.

\begin{figure}
    \centering
    \includegraphics[width=\columnwidth]{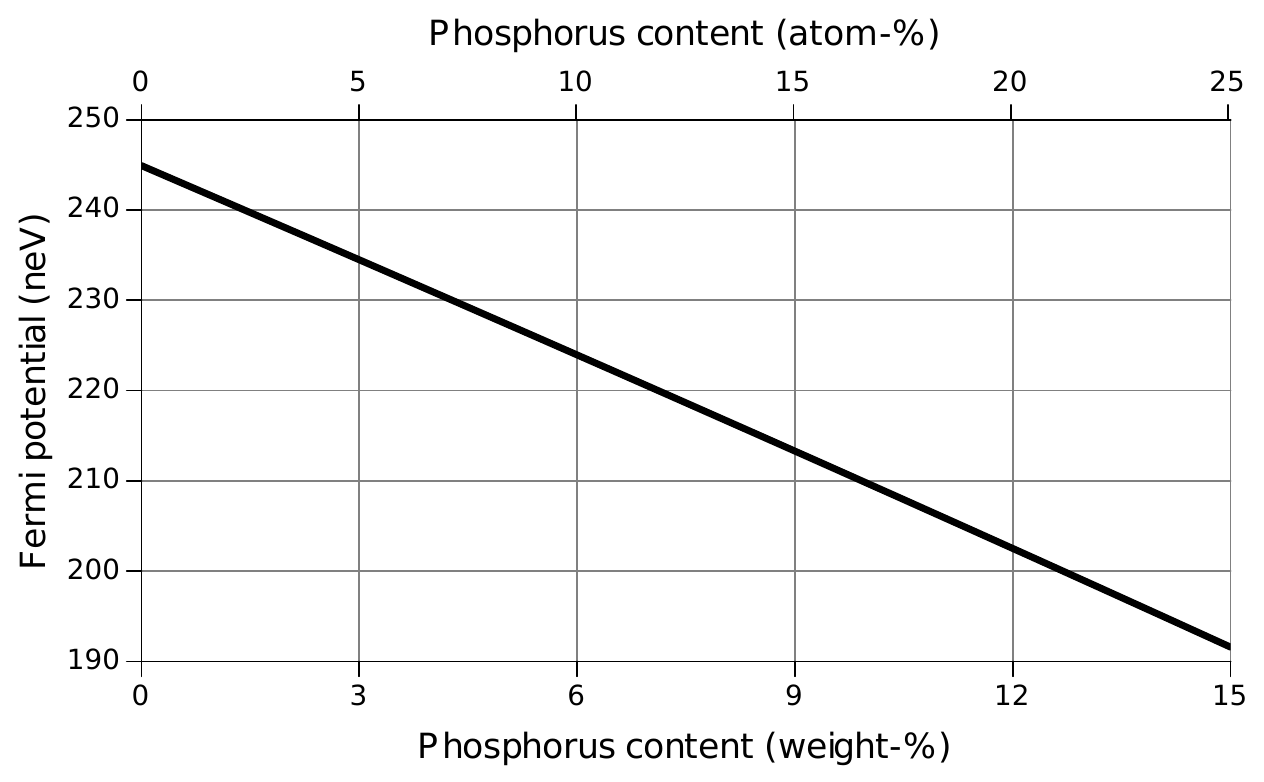}
    \caption{The Fermi potential of electroless nickel plating, calculated using neutron scattering lengths from~\cite{doi:10.1080/10448639208218770} and densities from~\cite{nickelInstitute}, drops with increasing phosphorus content.}
    \label{fig:NiP_Fermi}
\end{figure}

Electroless nickel plating is a coating process that can deposit a nickel-phosphorus mixture onto a variety of substrates, including steel, aluminum, and copper, by immersing them into a bath. It is a well-studied and widely established industrial process used to increase hardness and wear resistance of surfaces and is offered by a large number of vendors at low cost. The phosphorus content can be varied from less than \SI{4}{weight\percent} (''low-phosphorus``) to \SI{14}{weight\percent}, varying its metallurgical properties~\cite{nickelInstitute} and Fermi potential, see Fig.~\ref{fig:NiP_Fermi}. The thickness of the coating can typically be tuned from \SI{5}{\micro\meter} to \SI{50}{\micro\meter} by simply adjusting the time for which the substrate is submersed in the bath. Hardness can be further increased with heat treatment after plating.

Electroless-nickel-plated guides and vessels have been successfully used in a prototype UCN source developed at the Research Center for Nuclear Physics, Osaka University~\cite{PhysRevLett.108.134801,PhysRevC.99.025503} and for an upgrade of the UCN source at Los Alamos National Laboratory~\cite{PATTIE201764}. Above a phosphorus content of \SI{10}{weight\percent} (''high-phosphorus``) the coating becomes non-magnetic, making it suitable for transporting polarized UCNs, as demonstrated by \cite{TANG201632}.

In an effort to find a vendor capable of plating the \SI{2.67}{\meter}-long aluminum converter vessel for the new TUCAN source and UCN guides for a future TUCAN electric dipole moment (EDM) experiment, we used a variety of methods to characterize polished aluminum and stainless-steel substrates plated with electroless nickel by several vendors. We measured their roughness with a profilometer and we measured UCN storage lifetimes at the UCN sources at TRIUMF, Los Alamos National Laboratory, and J-PARC.

Based on the findings we selected a vendor to plate the converter vessel for the new TUCAN source and tested its UCN-storage properties at Los Alamos National Laboratory.

\section{Sample preparation}

\begin{figure}
    \centering
    \includegraphics[width=0.8\columnwidth,trim={7.5cm 8.5cm 7.4cm 3cm},clip]{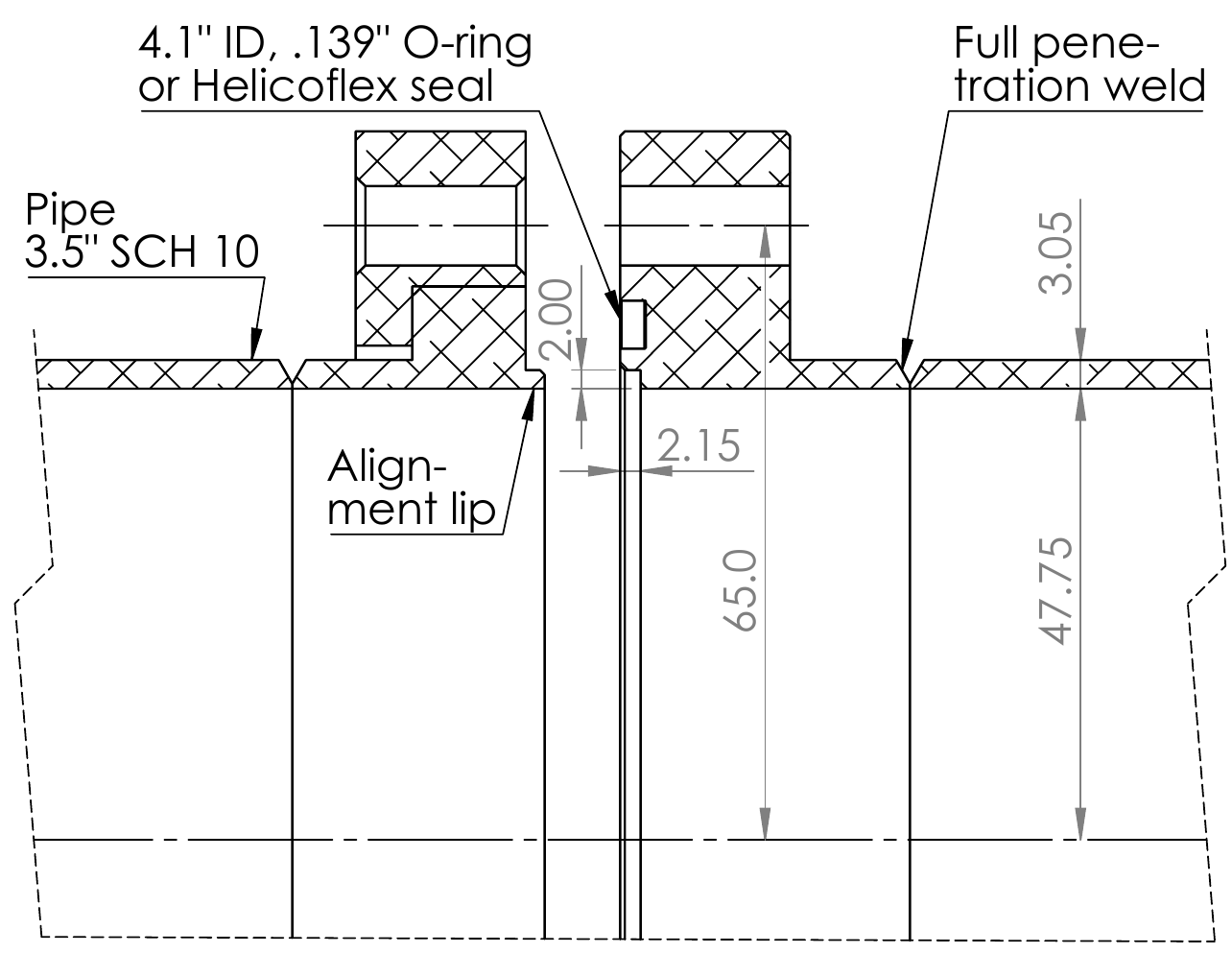}
    \caption{Design of UCN guides for the TUCAN source. All dimensions are given in millimeters.}
    \label{fig:UCNguide}
\end{figure}

In UCN-transport simulations of the new TUCAN source and EDM experiment we varied the guide diameters between \SIlist{85;125}{\milli\meter} and achieved the best transport efficiency with guide diameters above \SI{95}{\milli\meter}~\cite{SidhuThesis}. Hence, we chose a \SI{3.5}{in} schedule 10 pipe standard with \SI{4}{in} outer diameter and \SI{0.12}{in} wall thickness, resulting in an inner diameter of \SI{95.5}{\milli\meter}. For this pipe size we developed a new flange compatible with O-ring, \textit{HELICOFLEX}, and diamond-shaped aluminum seals. An alignment lip ensures that guides are connected seamlessly and concentrically, see Fig.~\ref{fig:UCNguide}.

\textit{Irving Polishing \& Manufacturing}\footnote{Irving Polishing \& Manufacturing, Inc. 5704-46th Street, Kenosha, WI 53144, USA, \url{irvinginc.com}} polished these guides to a nominal roughness $Ra$ (see section \ref{sec:profilometer}) of \SIrange{51}{76}{\nano\meter} (\SIrange{2}{3}{\micro in}), with the exception of one guide, which was polished to \SI{760}{\nano\meter} (\SI{30}{\micro in}).

We sent guides to three North American vendors for electroless nickel plating: \textit{Chem Processing}\footnote{Chem Processing, Inc. 3910 Linden Oaks Drive, Rockford, IL 61109, USA, \url{chemprocessing.com}} \textit{Advanced Surface Technologies}\footnote{Advanced Surface Technologies, Inc. 6155 W 5th Ave, Arvada, CO 80002, USA, \url{astfinishing.com}}, and \textit{Dav-Tech Plating}\footnote{Dav-Tech Plating, Inc. 40 Cedar Hill St, Marlborough, MA 01752, USA, \url{dav-techplatinginc.com}}. \textit{Chem Processing}, the same vendor used successfully for UCN guides by \cite{PATTIE201764,TANG201632}, plated three \SI{1}{\meter}-long guides, including the one with increased roughness. \textit{Advanced Surface Technologies} plated three pairs of \SI{0.5}{\meter}-long guides, including one pair with ''black electroless nickel``---a coating that we hoped would provide similar UCN-storage properties but a lower reflectivity for thermal radiation~\cite{B204483H}. \textit{Dav-Tech Plating} plated one \SI{1}{\meter}-long guide.

An additional \SI{1}{\meter}-long stainless steel guide was manufactured in Japan, plated by \textit{Asahi Precision}\footnote{Asahi Precision Co., Ltd. 30-5 Kaide-Cho, Mukou, Kyoto 617-0004, Japan, \url{akg.jp/precision}}, and finally polished to a nominal roughness of \SI{10}{\nano\meter} by \textit{Ultra Finish Technology}\footnote{Ultra Finish Technology Co., Ltd. 1-1-1 Heisei-Cho, Yokosuka, Kanagawa 238-0013, Japan, \url{uft.co.jp}}.

All platings were specified to be \SI{5}{\micro\meter} thick for aluminum substrates and \SI{50}{\micro\meter} thick for stainless steel substrates, with a high phosphorus content of at least \SI{10}{weight\%}. The thicker plating on slightly magnetic stainless steel should reduce the UCN-depolarization probability~\cite{TANG201632}.

Before any experiments with UCNs, we manually wiped the guides with a lukewarm \SI{1}{\percent} solution of \textit{Alconox} degreaser, de-ionized water, and isopropyl alcohol using \textit{AlphaWipe} cleanroom cloths. To replace a guide in the experimental setup we vented the vacuum with nitrogen or argon gas, replaced the respective UCN guide, and then evacuated the system again to a pressure below \SI{3e-5}{\milli\bar}.

The converter vessel for the new TUCAN source consists of two aluminum hemispheres with an inner radius of \SI{180}{\mm} and a \SI{2250}{\mm}-long tube with an inner diameter of \SI{148.2}{\mm}, welded into one \SI{2670}{\mm}-long vessel with only one open end, see also Fig.~\ref{fig:LANL}. While the tube was again polished by \textit{Irving Polishing \& Manufacturing} to a nominal $Ra$ of \SI{51}{\nano\meter} (\SI{2}{\micro in}), the hemispheres were only electropolished, with the roughest areas receiving some additional manual polishing to bring the roughness below \SI{1020}{\nano\meter} (\SI{40}{\micro in}).

The size of the parts to be plated are limited by the sizes of each vendor's plating baths, which were $\sim$\SI{1}{\meter} for \textit{Chem Processing} and $\sim$\SI{0.5}{\meter} for \textit{Advanced Surface Technologies}. \textit{Dav-Tech} was the only vendor with baths large enough to plate the challenging geometry of the converter vessel, requiring some additional plumbing to maintain a steady flow of plating solution through the vessel. After plating, we cleaned the vessel by filling it with a \SI{1}{\percent} solution of Alconox in de-ionized water and suspending an ultrasonic rod transducer into the solution for \SIrange{20}{40}{\minute}. We repeated the ultrasonic cleaning with only de-ionized water, flushing and rinsing the vessel with de-ionized water before and after. Due to the size and shape of the vessel, wiping the inner surfaces with isopropyl alcohol was not possible. Unfortunately, the de-ionized water left more residue as expected and an improved cleaning method will be needed in the future.

\section{Measurement techniques}
\label{sec:techniques}

\subsection{Profilometer and $Ra$ roughness}
\label{sec:profilometer}

We used a \textit{Mitutoyo Surftest SJ-210} profilometer to measure the roughness of all guides at several positions along their length and circumference. The profilometer drags a diamond stylus across the surface to measure its roughness amplitude and can provide a set of different measures of surface roughness. We generally used the most common, standardized $Ra$ measure to be able to compare to measurements done by the polishing companies. Localized scratches and imperfections in the plating can cause large outliers in roughness measurements, which we excluded from the data. We regularly checked the profilometer against a calibration sample with well-defined $Ra$ and also confirmed that it can measure $Ra$ roughnesses as small as \SI{2}{\nano\meter} on a glass tube.

\subsection{Storage of ultracold neutrons}

To compare storage properties with varying experimental setups at several UCN sources we used a detailed analytical model taking into account the gravitational potential $mgz$ along the elevation $z$ and the neutron's beta-decay lifetime $\tau_\beta$. The loss rate $\tau^{-1}$ for UCN with total energy $H = E + mgz$ can be calculated as
\begin{equation}
\label{eq:lossrate}
\tau^{-1}(H) = \frac{\sqrt{2H/m}}{4 \gamma(H)} \int_a^b \frac{H - mgz}{H} \sum_i \frac{dA_i}{dz} \mu_i(H - mgz) dz + \tau_\beta^{-1},
\end{equation}
where $a$ and $b$ are the minimum and maximum elevation within the storage vessel, and
\begin{equation}
\gamma(H) = \int_a^b \frac{dV}{dz} \sqrt{\frac{H - mgz}{H}} dz
\end{equation}
is the phase space available to UCN with total energy $H$. $\frac{dA_i}{dz}$ is the differential surface area covered by material $i$ and $\frac{dV}{dz}$ the differential volume of the storage vessel at elevation z. Refer to \cite[ch. 4.3]{golub_ultra-cold_1991} for a detailed explanation of this model.

The loss probability $\mu$ is calculated assuming an isotropic velocity distribution of UCN impinging on the surface with Fermi potential $U - iW$:
\begin{equation}
\begin{split}
\mu(E) = 
2 \int_0^{\pi/2} &\left( 1 - \left| \frac{\sqrt{E \cos^2 \theta} - \sqrt{E \cos^2 \theta - U + iW}}{\sqrt{E \cos^2 \theta} + \sqrt{E \cos^2 \theta - U + iW}} \right|^2 \right) \\
&\cdot \cos \theta \sin \theta d\theta.
\end{split}
\end{equation}

Assuming a spectrum of UCNs $\frac{dN}{dH}$ initially filled into the storage vessel the number of remaining UCNs $N(t)$ at time $t$ is then given by
\begin{equation}
N(t) = \int_{H_\mathrm{min}}^{H_\mathrm{max}} \frac{dN}{dH} \exp \left[ -t \cdot \tau^{-1}(H) \right] dH.
\label{eq:detailedStorage}
\end{equation}
Given a measurement of UCNs remaining in the vessel after different storage times we can fit the resulting curve with this model and determine the imaginary Fermi potential $W$. To compensate for detector noise and fluctuations in the source intensity, we determine the detector background rate and, where possible, estimate the UCN density delivered by the source using an independent normalization count $N_n$ during each filling of the storage vessel. The detector count $N_d(t)$ is then corrected by subtracting the expected background counts $N_b$ and dividing it by a normalization factor $N_n/\bar{N_n}$, where $\bar{N_n}$ is an average normalization count
\begin{equation}
    N(t) = \frac{N_d(t) - N_b}{N_n} \bar{N_n}.
\end{equation}

The energy spectrum provided by UCN sources is usually not very well known and is strongly shaped by guide geometry, guide materials, and energy-dependent detector efficiencies. Typically, it can only be indirectly inferred from simulations, which themselves have to make certain assumptions for these energy-dependent processes. Hence, we try to rely on simulations as little as possible and instead estimate parameterized spectra from simple principles while applying large uncertainties.

We include such systematic uncertainties by performing two fits for each model parameter, with its value set to the low end or the high end of its range while the other parameters are fixed to their central value. We then quote the mean $W$ of the resulting set of results $W_i$ and calculate its uncertainty as the square root of the quadratic sum of differences from the mean $\Delta W = \sqrt{ \sum \left( W_i - W \right)^2/2 }$. The factor $1/2$ in the sum accounts for the fact that we perform the fit twice for each parameter.

If the height of the storage vessel is small we can assume $\frac{dV}{dz} = V \delta({z - \bar{z}})$ and $\frac{dA_i}{dz} = A_i \delta({z - \bar{z}})$, simplifying the model for the loss rate to
\begin{equation}
\label{eq:simplelossrate}
\tau^{-1}(E) = \sqrt{\frac{2E}{m}} \sum_i \frac{A_i}{4 V} \mu_i(E) + \tau_\beta^{-1}
\end{equation}
where $E = H - mg \bar{z}$ is the kinetic energy at the center $\bar{z}$ of the storage vessel.

In simple storage experiments the energy-dependence of the loss rate is often ignored and instead an average storage lifetime $\bar{\tau}$ is determined by fitting a simple exponential model
\begin{equation}
N(t) = N_0 \exp \left( -t/\bar{\tau} \right).
\label{eq:storageLifetime}
\end{equation}
We list the result of such a fit for each measurement as well.

We calculated the real Fermi potential $U$ from neutron scattering lengths~\cite{doi:10.1080/10448639208218770} and densities~\cite{nickelInstitute} for electroless nickel (EN) and stainless steel (SS), giving $U_\mathrm{EN} = \SI{213+-5}{\nano\electronvolt}$ and $U_\mathrm{SS} = \SI{183+-5}{\nano\electronvolt}$, with uncertainties of $\SI{5}{\nano\electronvolt}$ to account for any under- or over-estimations. This is also in good agreement with measurements by \cite{PATTIE201764}, which gave potentials for electroless nickel of \SIrange{211+-5}{214+-5}{\nano\electronvolt}.

The imaginary Fermi potential correctly describes UCN losses if the loss cross section is indirectly proportional to the neutron velocity, which is the case for e.g.~the nuclear-absorption cross section. From catalogued absorption cross sections~\cite{doi:10.1080/10448639208218770} we expect imaginary Fermi potentials of about \SI{0.023}{\nano\electronvolt} for a high-phosphorus electroless nickel mixture and \SI{0.019}{\nano\electronvolt} for stainless steel. However, other loss processes, e.g.~UCNs escaping through gaps, may be better described with an energy-independent loss probability $\mu(E) = \bar{\mu}$.

\begin{table*}
    \centering
    \caption{Assumed energy range $H_\mathrm{min}$ to $H_\mathrm{max}$ and spectral exponent $\alpha$ of UCNs filled into each UCN-storage setup with guide elevation $\bar{z}$, stainless steel surface area $A_\mathrm{SS}$, electroless-nickel-plated surface area $A_\mathrm{EN}$, and volume $V$; including setups with the new TUCAN source converter vessel (CV).}
    \begin{tabular}{lrrrrrrr}
    \toprule
        Setup & $H_\mathrm{min}$ (neV) & $H_\mathrm{max}$ (neV) & $\alpha$ & $\bar{z}$ (cm) & $A_\mathrm{SS}$ (\si{\cm\squared}) & $A_\mathrm{EN}$ (\si{\cm\squared}) & $V$ (\si{\cubic\cm}) \\
        \midrule
        TRIUMF baseline & \num{10 +- 10} & \num{92 +- 10} & \num{0+-0.5} & -84 & 1109 & 0 & 2049 \\
        TRIUMF & \num{10 +- 10} & \num{92 +- 10} & \num{0+-0.5} & -84 & 1109 & 7476 & 9525 \\
        LANL1 high & \num{10 +- 10} & \num{102 +- 10} & \num{0.95+-0.5} & 0 & 387 & 3073 & 7838 \\
        LANL1 low & \num{10 +- 10} & \num{102 +- 10} & \num{0.95+-0.5} & -84 & 387 & 3073 & 7838 \\
        LANL2 high & \num{10 +- 10} & \num{102 +- 10} & \num{0.95+-0.5} & -24 & 315 & 3226 & 8007 \\
        LANL2 low & \num{10 +- 10} & \num{102 +- 10} & \num{0.95+-0.5} & -108 & 315 & 3226 & 8007 \\
        LANL2 high (CV) & \num{10 +- 10} & \num{102 +- 10} & \num{0.95+-0.5} & -24 & 473 & 14956 & 67643 \\
        LANL2 low (CV) & \num{10 +- 10} & \num{102 +- 10} & \num{0.95+-0.5} & -108 & 473 & 14956 & 67643 \\
        J-PARC & \num{64(10)} & \num{310(10)} & \num{0.75 +- 0.75} & 13 & 315 & 3226 & 8007 \\
        \bottomrule
    \end{tabular}
    \label{tab:setups}
\end{table*}

\subsubsection{Storage of ultracold neutrons at TRIUMF}
\label{sec:TRIUMFstorage}

\begin{figure}
    \centering
    \includegraphics[width=\columnwidth, trim=4.5cm 6.3cm 3cm 1.5cm, clip]{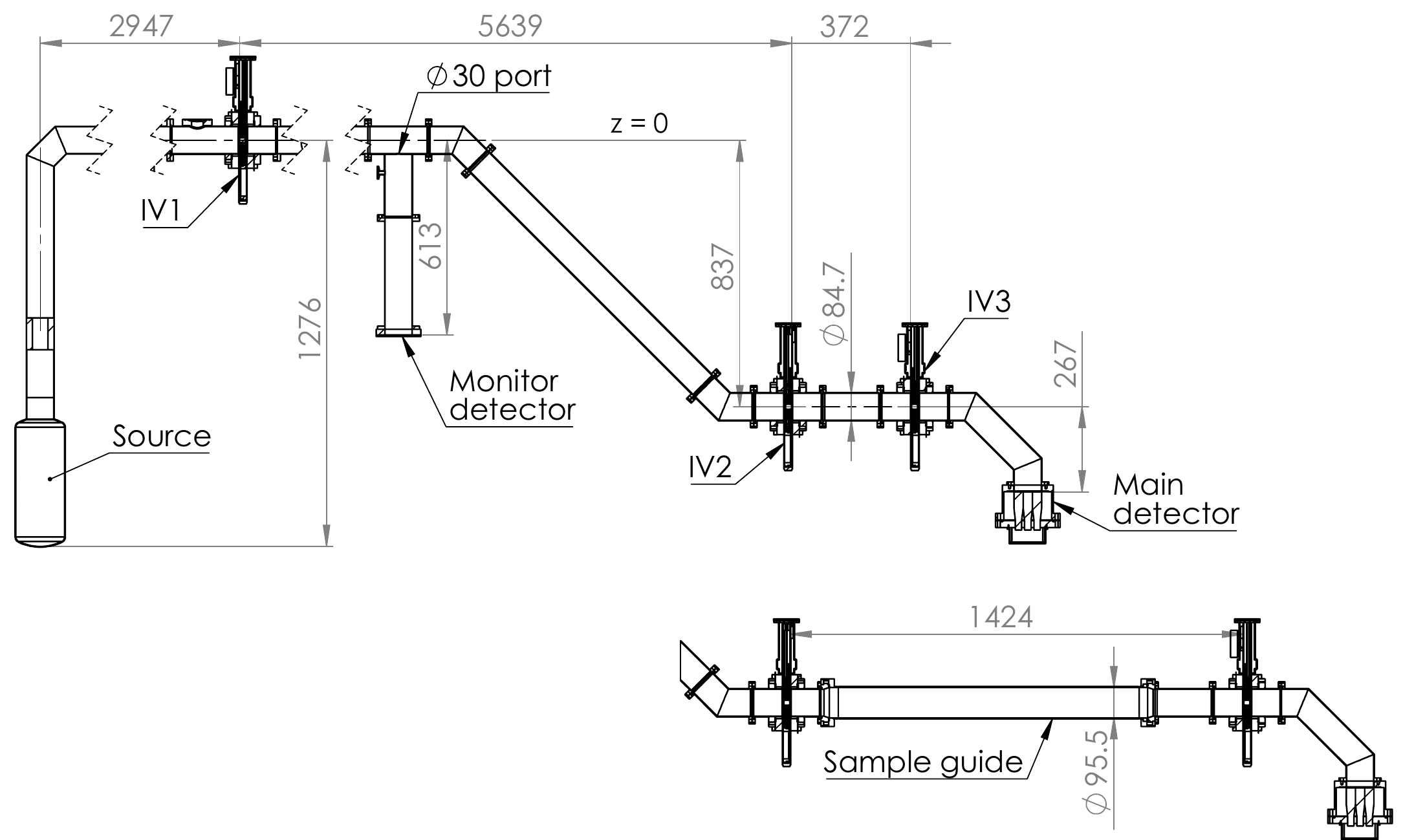}
    \caption{Side view of experimental setups to measure UCN storage lifetime using a prototype UCN source at TRIUMF. \textit{Top}: baseline; \textit{bottom}: with sample guide inserted. All dimensions are given in millimeters.}
    \label{fig:TCN19}
\end{figure}

To measure UCN storage lifetimes in guides at TRIUMF we used a prototype UCN source shown in Fig.~\ref{fig:TCN19}. UCNs are extracted vertically from a superfluid-helium converter and transported through stainless steel guides penetrating the radiation shielding, see also~\cite{PhysRevC.99.025503}. Where they exit the shielding, a monitor detector, a Dunia-10 $^3$He proportional counter with an aluminum window, is attached to the guide through a \SI{30}{\milli\meter}-diameter port. The main guide continues downwards \SI{84}{\centi\meter} to the storage chamber.

Three valves---IV1, IV2, and IV3---are mounted in the beamline. To fill the storage chamber, we opened IV1 and IV2 and irradiated the neutron spallation target with protons to produce ultracold neutrons. After a filling period of \SI{60}{\second} we closed IV1 and IV2 and turned off the proton beam. UCNs are now stored between IV2 and IV3. After a storage period we opened IV3 and dumped the remaining UCNs into the main detector, a $^6$Li-glass-based scintillation detector~\cite{Jamieson2017}. We varied the storage period from \SIrange{2}{150}{\second} to observe the exponential drop in the number of detected UCNs and determine the storage lifetime. The number of UCNs detected in the monitor detector during the storage and counting periods serves as normalization. During the storage periods we determined the average background rate in the main detector.

The valves are off-the-shelf \textit{VAT 17.2 series} gate valves with a protective ring improving UCN transmission in the open state. We modified them with a shim plate to eliminate gaps between the vacuum-sealing blade and the attached guides, drastically improving their storage properties in the closed state~\cite{dstang}.

Since we performed all measurements at the same elevation and the elevation difference within the sample guides is small we used the simplified equation~(\ref{eq:simplelossrate}) for the loss rate $\tau^{-1}(E)$. To estimate the energies of UCNs in the storage chamber, we assumed that the maximum energy of UCNs produced in the source is \SI{213}{\nano\electronvolt}, the Fermi potential of its electroless nickel plating. To exit the source the UCNs have to overcome an elevation difference from the bottom of the source vessel to the exit guide corresponding to a gravitational potential of \SI{131}{\nano\electronvolt}. The resulting spectrum of UCNs exiting the source is relatively narrow with energies between $H_\mathrm{min} = \SI{0}{\nano\electronvolt} + \Delta H$ and $H_\mathrm{max} = \SI{82}{\nano\electronvolt} + \Delta H$. Since energy cutoffs are typically not perfectly sharp, we added offsets $\Delta H = \SI[separate-uncertainty=true]{10+-10}{\nano\electronvolt}$ with a large uncertainty.

\begin{figure}
    \centering
    \includegraphics[width=\columnwidth]{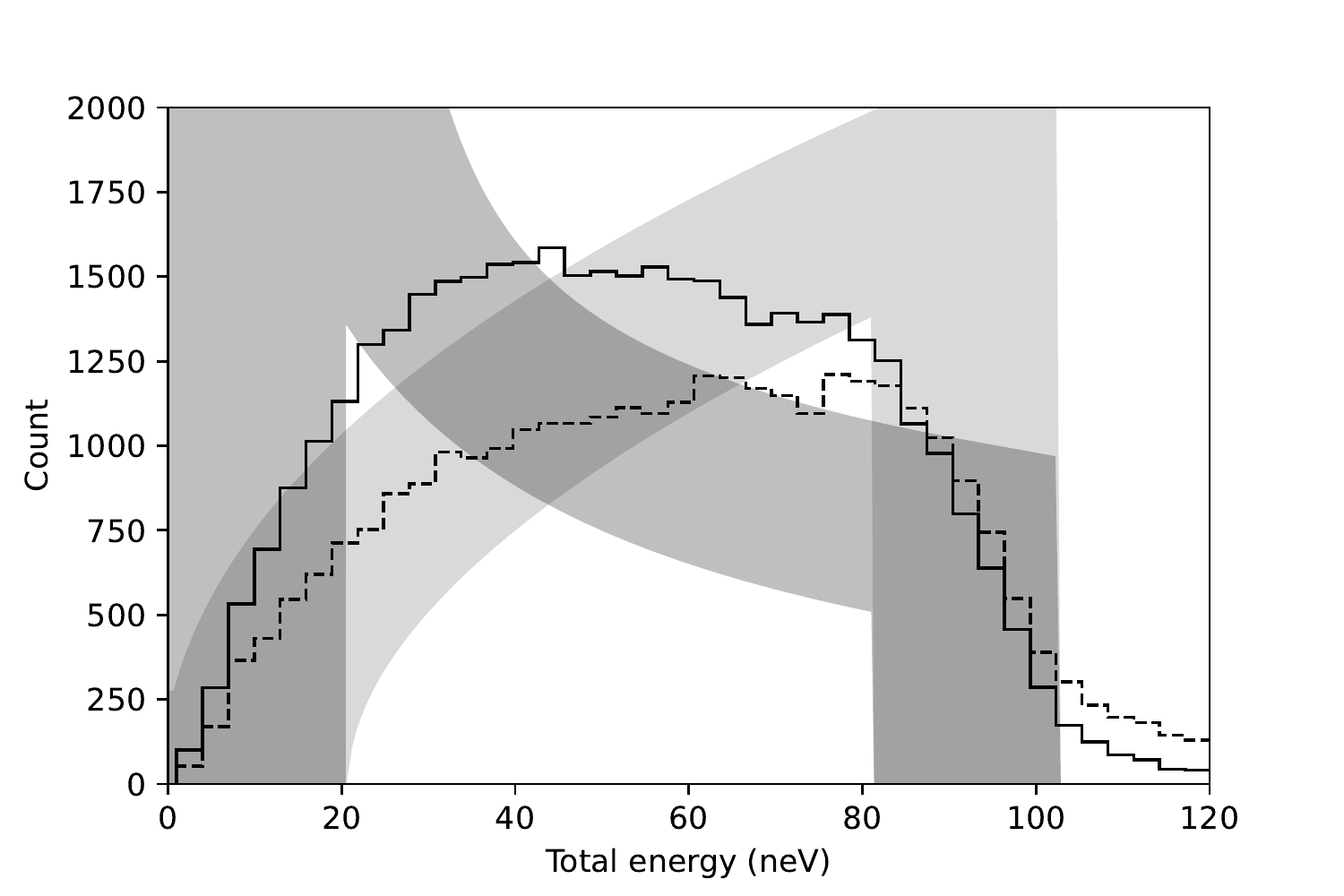}
    \caption{Simulated total-energy spectrum of UCN detected after a \SI{0}{\second}-long storage in the TRIUMF setup, assuming a $\sqrt{E}$ spectrum generated in the source and surfaces with imaginary Fermi potentials (solid line) and with energy-independent loss per bounce (dashed line). The shaded areas indicate the extremes of the assumed spectrum shape in the analytical model with $\alpha$ between -0.5 and 0.5 and $\Delta H = \SI{10}{\nano\electronvolt}$.}
    \label{fig:Hend}
\end{figure}

Simulations of this source with PENTrack~\cite{SCHREYER2017123,PhysRevC.99.025503} indicate that the stored spectrum is relatively flat, so we assumed an energy spectrum filled into the storage vessel of the form
\begin{equation}
\frac{dN}{dH} \propto \left(H - H_\mathrm{min} \right)^\alpha
\label{eq:spectrum}
\end{equation}
with an exponent $\alpha = \num[separate-uncertainty=true]{0 +- 0.5}$, again with a large uncertainty to account for any neglected effects that might modify the spectrum, see Fig.~\ref{fig:Hend}.

The valves and their adapters are bare stainless steel. We were able to measure its imaginary Fermi potential $W_\mathrm{SS}$ in a baseline experiment with only stainless steel surfaces, see Fig.~\ref{fig:TCN19}. After adding an electroless-nickel-plated sample guide with two short adapters\footnote{The adapters are polished aluminum plated by Chem Processing and \SI{24}{\milli\meter} long.} we were able to subtract the losses on the stainless steel surfaces and calculate the imaginary Fermi potential $W_\mathrm{EN}$ of only the electroless nickel plating.

All relevant parameters---assumed energy spectra, elevations, surface areas, and volumes---are listed in table~\ref{tab:setups}.

\subsubsection{Storage of ultracold neutrons at Los Alamos National Laboratory}

\begin{figure}
    \centering
    \includegraphics[width=\columnwidth, trim=3cm 9cm 2cm 4cm, clip]{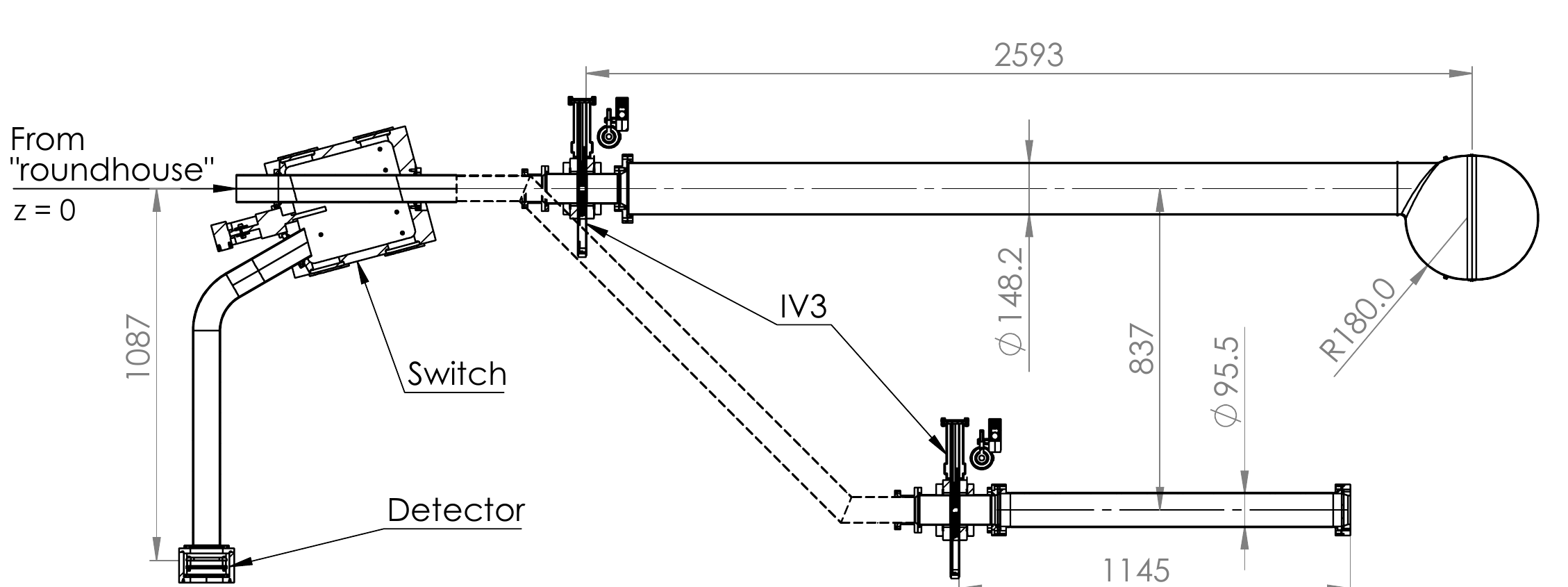}
    \caption{Side view of experimental setups to measure UCN storage lifetime at Los Alamos National Laboratory (``LANL1''), showing the TUCAN source converter vessel in the ``high'' position and a sample guide in the ``low'' position. A second set of measurements (``LANL2'') was performed with a different switch that added an additional drop of \SI{24}{\cm} from the roundhouse. All dimensions are given in millimeters.}
    \label{fig:LANL}
\end{figure}

We also performed measurements with two sample guides and eventually the converter vessel for the new TUCAN source at the UCN source at Los Alamos National Laboratory (LANL). At the LANL source, UCN are produced in a solid-deuterium crystal and then vertically extracted. Previous studies of this source suggest that a connected storage volume will be filled with a velocity spectrum $dN/dv \sim v^{2.9}$~\cite{PhysRevC.97.012501}, corresponding to an energy spectrum\footnote{For a velocity spectrum $dN/dv \sim v^\beta$ the energy spectrum becomes $\frac{dN}{dE} = \frac{dN}{dv}\frac{dv}{dE} \sim v^\beta E^{-1/2} \sim E^{\beta/2 - 1/2}$.} following equation~(\ref{eq:spectrum}) with $\alpha = \num[separate-uncertainty=true]{0.95+-0.5}$, where we again assumed a large uncertainty on the exponent. We performed our measurements at a beam port supplied through the "roundhouse", a large storage volume containing a UCN absorber at a height of \SI{90}{\cm}, limiting the UCN spectrum to energies between $H_\mathrm{min} = \SI{0}{\nano\electronvolt} + \Delta H$ and $H_\mathrm{max} = \SI{92}{\nano\electronvolt} + \Delta H$. Again, we added large uncertainties $\Delta H = \SI[separate-uncertainty=true]{10+-10}{\nano\electronvolt}$, in part to take into account that UCNs with energies only slightly above the absorber threshold may take a long time to be absorbed.

The storage volume was filled for \SI{200}{\second} from the source, through a superconducting polarizer magnet, the roundhouse, a switch, and a storage valve, see Fig.~\ref{fig:LANL}. Then, we closed the storage valve and moved the switch to dump UCNs remaining in the guides into the $^{10}$B-coated ZnS:Ag-scintillation detector~\cite{WANG2011323,Wong22}; this count served as normalization. After a storage period of at least \SI{20}{\s}---the minimum time needed to completely empty the guides---we opened the valve again to count the UCNs remaining in the storage volume. During the following waiting period we measured the average background rate in the detector.

We performed the measurements at two elevations, a high position with switch and storage valve at the same level, and a low position with a vertical drop of \SI{84}{\cm}. For a second set of measurements including the converter vessel we used a different switch with an additional drop of \SI{24}{\cm} between roundhouse and switch.

To take into account the elevation changes and the larger elevation difference in the converter vessel we used the full equation (\ref{eq:lossrate}) for the loss rate $\tau^{-1}(H)$. We assumed that the small portion of stainless steel surfaces have the same Fermi potential as was determined in the TRIUMF baseline measurement. All relevant parameters are listed in table~\ref{tab:setups}.

\subsection{Storage of ultracold neutrons at J-PARC}

\begin{figure}
    \centering
    \includegraphics[width=\columnwidth, trim=1.5cm 7cm 1.5cm 2.5cm, clip]{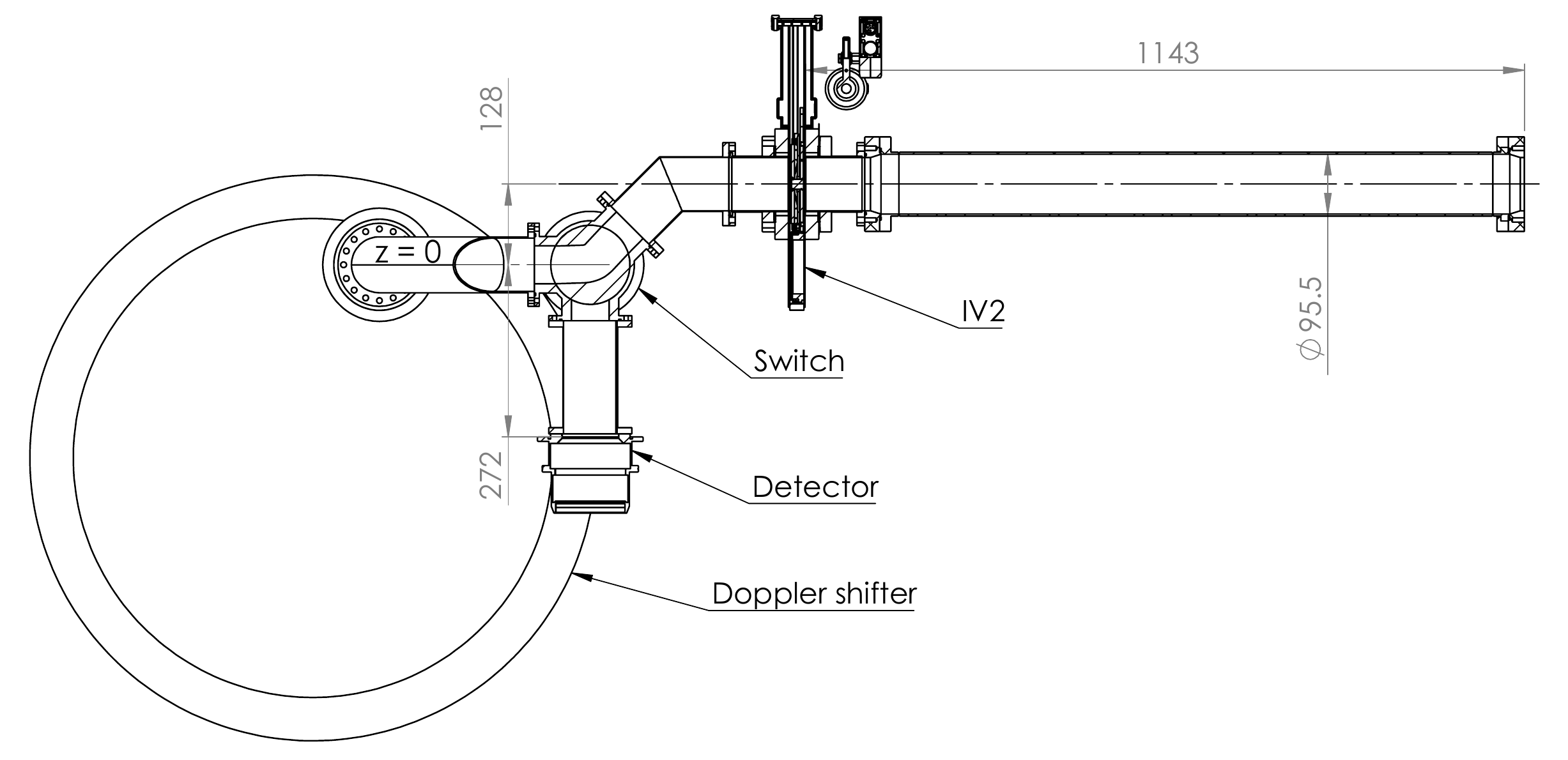}
    \caption{Side view of experimental setup to measure UCN storage lifetime at J-PARC. All dimensions are given in millimeters.}
    \label{fig:JPARC}
\end{figure}

We performed another measurement at a UCN source at J-PARC, which uses a mirror mounted on a rotating arm to Doppler-shift a cold-neutron beam~\cite{10.1093/ptep/ptv177}. We extracted the resulting ultracold neutrons through an aluminum foil and filled them for \SI{40}{\second} through a switch with a \SI{13}{\centi\meter} upward step and through a valve into the storage volume, see Fig.~\ref{fig:JPARC}. Then we closed the valve to store UCNs while we moved the switch towards the detector and removed the UCNs remaining in the guides. After a storage period of \SIrange{25}{125}{\second} we opened the valve again and counted the remaining UCNs with a Dunia-10 $^3$He proportional counter with aluminum window. During the waiting periods before and after counting we determined the average background rate in the detector.

Although we measured time-of-flight spectra of UCNs leaving the Doppler shifter, showing a flux strongly peaked at a velocity of \SI{7.5}{\meter\per\second}, the spectrum filled into the storage volume is difficult to estimate. Hence, we assumed the spectrum follows equation~(\ref{eq:spectrum}) between $H_\mathrm{min} = \SI{54}{\nano\electronvolt} + \Delta H$, the energy required to overcome the aluminum foil, and $H_\mathrm{max} = \SI{300}{\nano\electronvolt} + \Delta H$ with potential offsets $\Delta H = \SI[separate-uncertainty=true]{10+-10}{\nano\electronvolt}$, and with an exponent $\alpha = \num[separate-uncertainty=true]{0.75+-0.75}$ with a very large uncertainty. Since our model for the loss rate takes into account the quick loss of any UCNs above the Fermi potential of the storage volume the obtained results are not sensitive to the value of $H_\mathrm{max}$. Since we only performed measurements with the small sample guides, we used the simplified equation~(\ref{eq:simplelossrate}) to determine the imaginary Fermi potential. For the stainless steel surfaces in the storage volume we again assumed their Fermi potential is the same as measured in the TRIUMF baseline experiment. All relevant parameters are listed in table~\ref{tab:setups}.

\section{Results}

\subsection{$Ra$ roughness}

\begin{table}
    \centering
    \caption{Measured roughnesses of UCN guides and converter vessel (CV) plated by \textit{Chem Processing} (Chem Proc.), \textit{Ultra Finish Technology} (UFT), \textit{Advanced Surface Technologies} (AST), or \textit{Dav-Tech Plating}.}
    \begin{tabular}{lrrrr}
    \toprule
          &         & Sub-   & \multicolumn{2}{c}{$Ra$ roughness (nm)} \\
        \# & Plating & strate & before plating & after plating \\
    \midrule
        1 & Chem Proc. & Al & 54--62 & 87--197 \\
        2 & Chem Proc. & Al & 98--100 & 138--296 \\
        3 & Chem Proc. & Al & 744--752 & 850--976 \\
        4 & UFT & SS & -- & 40--105 \\
        5 & AST & Al & 52--78 & 83--237 \\
        6 & AST & SS & 63--76 & 63--130 \\
        7 & AST (black) & SS & 62--76 & 105--1326 \\
        8 & Dav-Tech & Al & 50--75 (nom.) & 254--737 \\
        CV & Dav-Tech & Al & 38--64$^\dagger$ (1040*) & 374--626$^\dagger$ \\
    \bottomrule
        \multicolumn{5}{r}{\footnotesize{$^\dagger$Tube section only. *Max. roughness on hemispheres}}
    \end{tabular}
\label{tab:guideroughness}
\end{table}

The electroless nickel plating provided by \textit{Chem Processing} and \textit{Advanced Surface Technologies} seems to typically increase the $Ra$ roughness and its variability by up to \SI{200}{\nano\meter}, independent of the substrate roughness, see table~\ref{tab:guideroughness}.

Although the roughness of the guide manufactured by \textit{Ultra Finish Technology} was considerably lower than the others, we measured values well above the quoted roughness of \SI{10}{\nano\meter}.

The ``black'' electroless nickel plating showed extreme variability in roughness.

The plating by \textit{Dav-Tech} also had larger variability. Some test tube samples showed a continuous variation along the inner circumference with the minimal and maximal roughness \SI{180}{\degree} apart. This possibly stems from bubbles created during the plating process rising in the bath, adhering to the upper half of the inner surface, and intermittently disrupting the plating process. \textit{Dav-Tech} was able to improve this variability by differently orienting some smaller samples in the plating bath.

Once the converter vessel was completed we were not able to measure roughness inside the hemispheres anymore, hence table~\ref{tab:guideroughness} lists post-plating roughness only for the tube section.

\subsection{Storage of ultracold neutrons}

\begin{figure}
    \centering
    \includegraphics[width=\columnwidth]{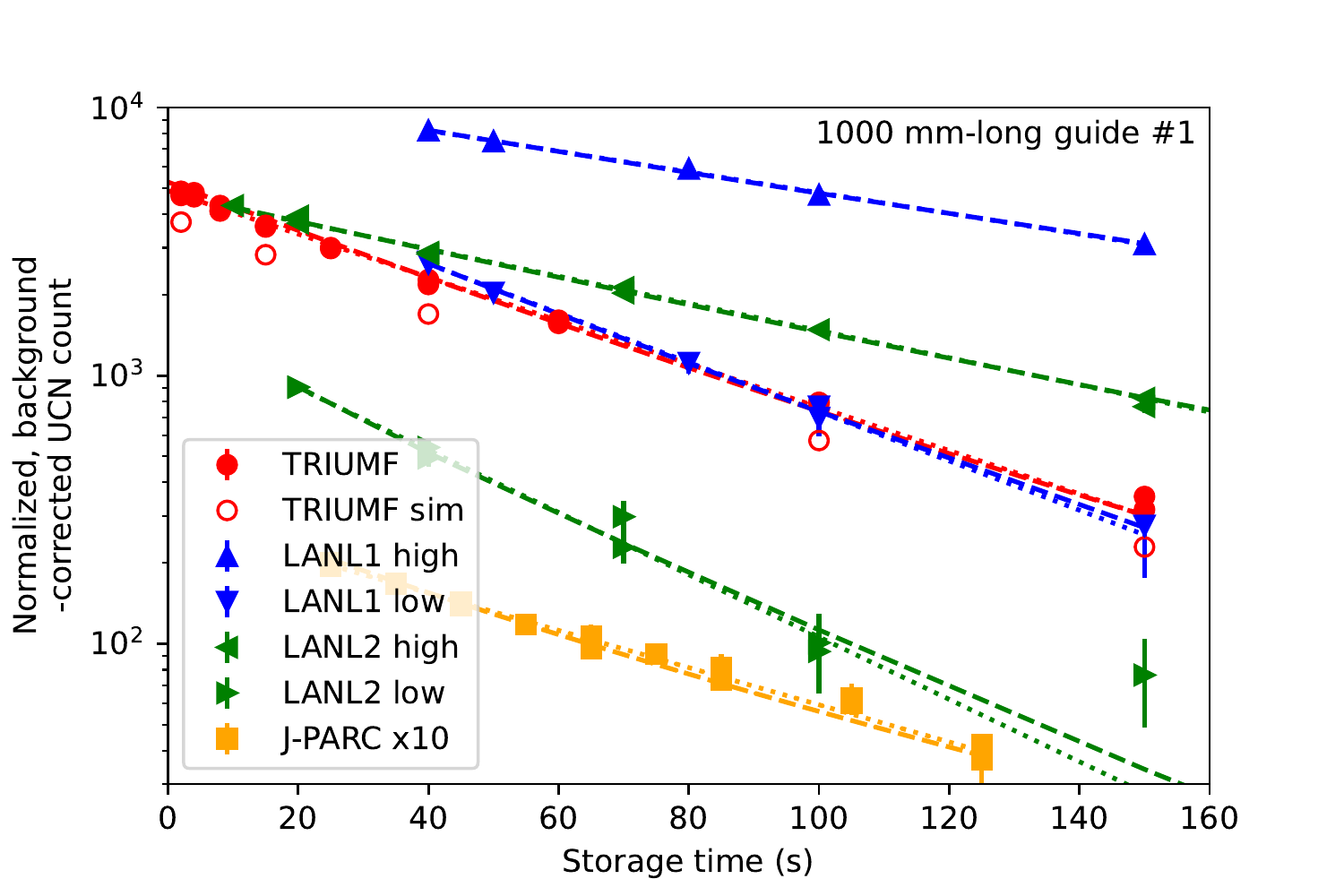}
    \includegraphics[width=\columnwidth]{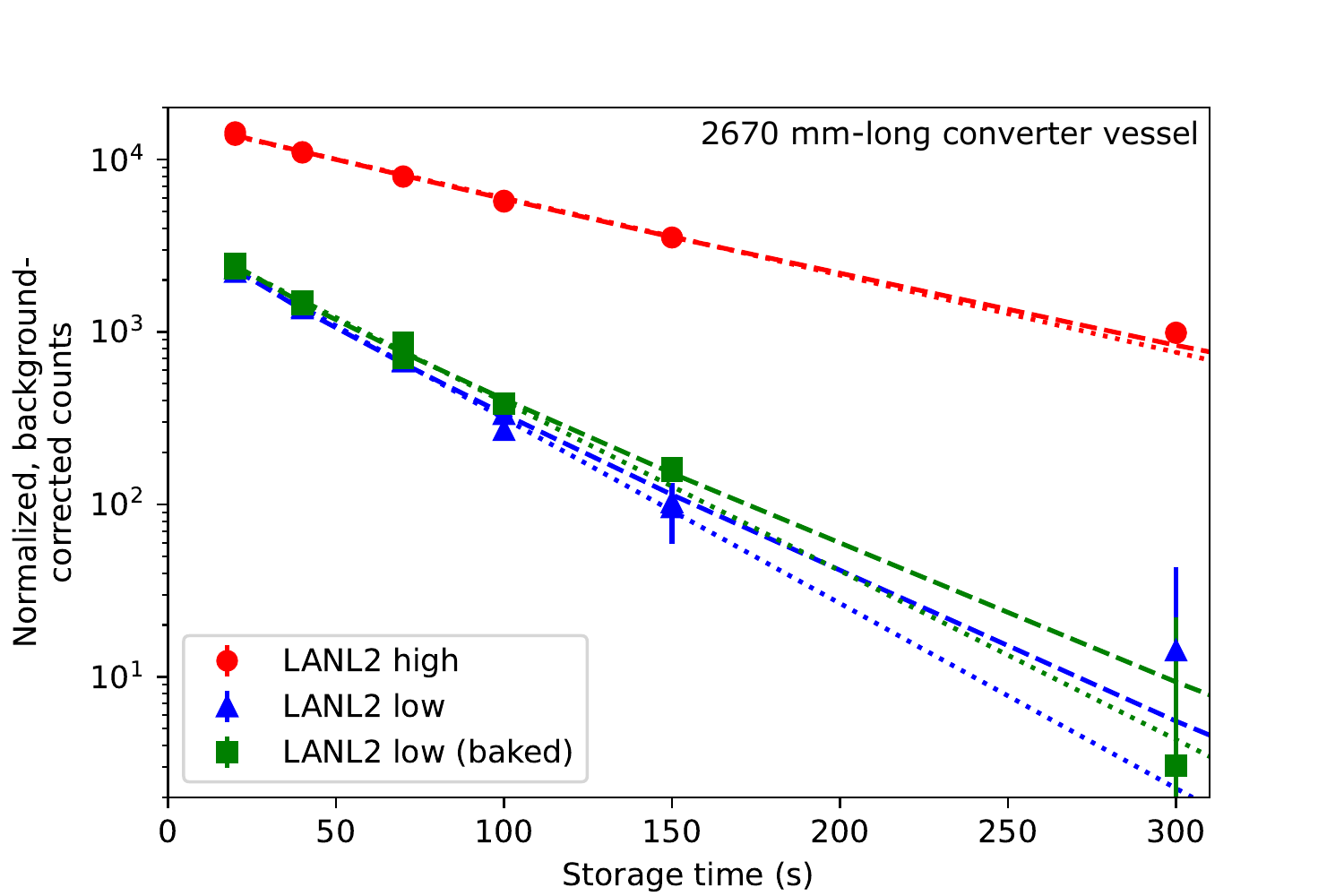}
    \caption{Number of UCNs counted after different storage times in reference guide \#1 (top) and in the converter vessel (bottom) after background subtraction and normalization. Error bars are calculated from statistical uncertainties ($\Delta N = \sqrt{N}$) of main, monitor, and normalization counters. Dashed and dotted lines show fits with the detailed model (\ref{eq:detailedStorage}) and with the average storage-lifetime model (\ref{eq:storageLifetime}). The results from the low-intensity J-PARC source are averaged over up to 80 fillings to reduce statistical uncertainties and are shown scaled by a factor of 10. Simulations of the TRIUMF setup (empty circles) with $W_\mathrm{EN} = \SI{0.04}{\nano\electronvolt}$ and $W_\mathrm{SS} = \SI{0.06}{\nano\electronvolt}$ replicate the slope in the experimental data well.}
    \label{fig:storageCurves}
\end{figure}

The UCN counts after different storage times in guide \#1 and in the converter vessel are shown in Fig.~\ref{fig:storageCurves}. With the limited statistics and relatively short storage times neither the detailed model (\ref{eq:detailedStorage}) nor the simple averaged storage-lifetime model (\ref{eq:storageLifetime}) provided a consistently better fit.

\begin{table}
    \centering
    \setlength{\tabcolsep}{3pt}
    \caption{List of storage setups (see table~\ref{tab:setups}) with guide \# (see table~\ref{tab:guideroughness}), their average storage lifetimes $\bar{\tau}$, energy-independent loss probabilities $\bar{\mu}$, and Fermi potentials $U - \mathrm{i}W$.}
        \begin{tabular}{lrrrrr}
    \toprule
        Setup & \# & $\bar{\tau}$ (s) & $\bar{\mu}$ (\num{e-4}) & $U - iW$ (neV) \\
    \midrule
        Baseline & -- & \num{31.4+-0.8} & \num{4.7(3)} & \num{183(5) - i 0.055(10)} \\
        TRIUMF & 1 & \num{53.8+-0.6} & \num{2.8(2)} & \num{213(5) - i 0.047(12)} \\
        TRIUMF & 2 & \num{53.1+-0.6} & \num{2.9(3)} & \num{213(5) - i 0.048(13)} \\
        TRIUMF & 3 & \num{49.8+-0.6} & \num{3.2(3)} & \num{213(5) - i 0.054(13)} \\
        TRIUMF & 4 & \num{41.8+-0.6} & \num{4.1(3)} & \num{213(5) - i 0.071(16)} \\
        TRIUMF & 5 & \num{42.7+-0.5} & \num{4.0(3)} & \num{213(5) - i 0.069(16)} \\
        TRIUMF & 6 & \num{47.2+-0.6} & \num{3.5(3)} & \num{213(5) - i 0.059(14)} \\
        TRIUMF & 7 & \num{12.7+-0.3} & \num{18.1(12)} & \num{213(5) - i 0.36(6)} \\
    \midrule
        LANL1 high & 1 & \num{112.5+-2.5} & \num{1.5(2)} & \num{213(5) - i 0.047(11)} \\
        LANL1 low & 1 & \num{47.3+-1.2} & \num{3.2(2)} & \num{213(5) - i 0.045(6)} \\
        LANL1 high & 8 & \num{127.4+-4.6} & \num{1.0(17)} & \num{213(5) - i 0.037(9)} \\
        LANL1 low & 8 & \num{45.4+-1.3} & \num{3.4(2)} & \num{213(5) - i 0.048(7)} \\
    \midrule
        LANL2 high & 1 & \num{85.7+-2.9} & \num{2.0(2)} & \num{213(5) - i 0.046(8)} \\
        LANL2 low & 1 & \num{37.4+-2.1} & \num{3.9(4)} & \num{213(5) - i 0.050(7)} \\
        LANL2 high & CV & \num{97+-3.0} & \num{3.4(2)} & \num{213(5) - i 0.072(11)} \\
        LANL2 low & CV & \num{40.5+-1} & \num{6.2(5)} & \num{213(5) - i 0.078(9)} \\
        LANL2 low* & CV & \num{44.4+-1.5} & \num{5.6(5)} & \num{213(5) - i 0.070(8)} \\
    \midrule
        J-PARC & 1 & \num{62.6 +- 1.8} & \num{2.7(3)} & \num{213(5) - i 0.049(16)} \\
    \bottomrule
        \multicolumn{5}{r}{\footnotesize{*After \SI{12}{\hour} baking at \SI{100}{\degreeCelsius}}}
    \end{tabular}
\label{tab:guidestorage}
\end{table}

The fit results for all measurements are shown in table~\ref{tab:guidestorage}. Despite the large systematic uncertainties in the energy spectra the estimated loss probabilities and imaginary Fermi potentials have uncertainties of less than \SI{25}{\percent}. The statistical uncertainties are small in comparison, even when scaled with the poor $\chi^2$ that is typical when fitting storage-lifetime measurements.

The imaginary Fermi potentials measured for guide \#1 in different setups are compatible, providing confidence that our model is well suited to compare measurements across several UCN sources. Especially measurements at varying elevations provide compatible imaginary Fermi potentials, indicating that it correctly describes the energy dependence of UCN losses. In contrast, the estimated energy-independent loss probabilities show large variations between different energy ranges, indicating that losses are dominated by absorption or upscattering and not by UCN escaping through gaps.

The imaginary Fermi potential of the TUCAN source converter vessel turned out to be \SI{50}{\percent} higher than the best guides. Baking the vessel at \SI{100}{\degreeCelsius} for \SI{12}{\hour} reduced the losses by \SI{10}{\percent}. Earlier tests with a smaller electroless-nickel-plated storage vessel at TRIUMF already showed that baking at a higher temperature of \SI{150}{\degreeCelsius} provides no further improvement.

We replicated the storage-lifetime measurements at TRIUMF with PENTrack simulations, see also~\cite{SCHREYER2017123,PhysRevC.99.025503}. The imaginary Fermi potentials we used in the simulations to match the measured lifetimes fell well within the estimated uncertainties, see Fig.~\ref{fig:storageCurves}.

\section{Conclusions}

We successfully compared UCN-storage measurements of electroless-nickel-plated guides performed at several UCN sources. Guides plated by \textit{Chem Processing} and \textit{Dav-Tech} achieved the best storage properties, with imaginary Fermi potentials between \SIlist{0.037;0.050}{\nano\electronvolt} or ratios of imaginary to real Fermi potential $\eta$ between \numlist{1.7e-4;2.3e-4}, a significant improvement over stainless steel with ${\eta = \num{3.0e-4}}$. \cite{PATTIE201764} reported similar values between \numlist{1.1e-4;2.7e-4}.

The ``black'' electroless nickel plating by \textit{Advanced Surface Technologies} unfortunately was not suitable for UCN storage, causing high losses and a four times shorter storage lifetime. However, diamond-like carbon may be an alternative for a UCN-reflecting coating~\cite{PhysRevC.74.055501} that absorbs thermal radiation~\cite{HANZELKA2008455}.

$Ra$ roughnesses as large as \SI{1000}{\nano\meter} seem to have little impact on the storage properties but larger roughness likely will negatively impact UCN transport. Although \textit{Chem Processing} was able to achieve improved roughness compared to \textit{Dav-Tech}, we had to choose \textit{Dav-Tech} to plate the converter vessel for the new TUCAN source, as they were the only vendor capable of plating such a big vessel.

The UCN losses in the plated converter vessel were measured to be \SI{50}{\percent} higher than in the best guides. However, simulations showed that that would have only a minor impact on the future TUCAN EDM experiment~\cite{SidhuThesis} since the losses in the source will be dominated by upscattering in the liquid helium converter. Hence, we consider the vessel acceptable for use in the new source. A closer inspection revealed that some residue was left after the de-ionized water rinse and we expect to be able to improve the storage properties with an improved cleaning procedure using an articulated arm to manually wipe the inner surface.

\section*{Acknowledgement}
We would like to thank C.~Marshall, S.~Horn, and D.~Rompen for engineering support; and B.~Hitti and C.~Dick for operational support of the helium liquefier at TRIUMF.

This research was supported by CFI project 36322, NSERC grant SAPPJ-201-00031, and JSPS KAKENHI grants 18H05230 and 20KK0069.
The measurement at LANL was supported by Los Alamos National Laboratory LDRD Program (Project No. 20190041DR).
The neutron experiment at the J-PARC Materials and Life Science Experimental Facility was performed under KEK S-type proposal 2019S03.




\bibliographystyle{elsarticle-num-names}
\bibliography{sample.bib}

\end{document}